\documentclass[english]{revtex4}
\usepackage{graphicx}
\usepackage{amsmath}
\usepackage{bm}
\usepackage[english]{babel}
\begin{document}

\title{Structural information extracted from the diffraction of XFEL fs-pulses in a crystal}

\author{A. Leonov$^1$, D. Ksenzov$^2$, A. Benediktovitch$^1$, I. Feranchuk$^1$ and U. Pietsch$^2$}

\address{$^1$Department of Theoretical Physics, Belarusian State University, 220030 Nezavisimosti ave. 4, Minsk, Belarus \\
$^2$ Festk\"{o}rperphysik, Universit\"{a}t Siegen, 57072 Walter-Flex-Str. 3, Siegen, Germany}

\begin{abstract}
\noindent We present a theoretical justification for a method of extracting of supplementary information for the phase retrieval procedure taken from diffraction of fs-pulses from X-ray Free Electron Laser facilities. The approach is based on numerical simulation of the dynamics of the electron density in the crystal composed of different atoms in the unit cell, namely a bi-atomic crystal containing heavy and light atoms. It is shown that evaluation of  diffraction intensities measured by means of different values of XFEL pulse parameters enables to find absolute values of structure factors for both types of atoms and their relative phase. The accuracy of structural information is discussed in terms of fluctuations of the evaluated atomic scattering factors. Our approach could be important for improvement of phase retrieval methods with respect to a more efficient determination of atomic positions within the unit cell of macromolecules.
\end{abstract}

\maketitle

PACS number(s): 32.80.Fb, 32.90.+a, 87.59..e, 87.15.ht

\section{Introduction}
\label{sec1:intro}

The solution of phase problem is one of the major tasks in X-ray structure analysis in order to determine the electron density distribution within the crystalline unit cell \cite{Taylor}. Numerous methods of direct and indirect solution of the phase problem exist but most of them are based on an initial guess of at least few of scattering phases and possible atomic positions of the studied unit cell \cite{Giac-Phase}. The phase problem becomes especially complicated in case of crystals with large number of atoms in the unit cell such as in macromolecules and proteins \cite{Chapman2011}. Moreover, some of such crystals are very difficult to produce and they are available only in small sizes ($\sim 10^3$ unit cells for the whole sample in 3D) that makes their investigation by means of conventional X-ray sources rather intricate.

X-ray Free Electron Lasers (XFEL) are often denoted as 4th generation X-ray sources providing extremely bright and ultrashort X-ray pulses that give unique possibilities to study the structure of matter with angstrom resolution on a time scale of femtoseconds \cite{FirstSLAC,FirstSACLA}. These unique properties of the XFEL pulses led to the development of some qualitatively new methods, where the serial femtosecond crystallography (SFX) is the most prominent one. A set of recent experiments \cite{Redecke-SFX} proved SFX to be a perfect tool for investigation of small crystals (nanocrystals) assuming that structure data can be taken before sample destruction takes place on a time scale larger than the FEL pulse length. Refinement of the diffraction data and solving the phase problem in most of these experiments was performed on the basis of the method of molecular replacement that still needs the implementation of already known very similar structure entities. However, the problem of direct measurement of structural amplitudes is one of the most challenging issues of using an XFEL for X-ray structural analysis \cite{Taylor}. Some theoretical investigations in this direction considering the XFEL-specific properties have already been published by Son et al. (the high-intense version of the MAD approach) \cite{Son-MAD} and Spence et al. (approach based on analysis of scattering intensities between Bragg reflections from nanocrystals) \cite{Spence-bB}. In the present paper we show that diffraction of XFEL femtosecond pulses from crystals could also provide extra information which can be used as initial constraints in the phase retrieval procedures which are not available in the conventional X-ray structure analysis.

The present paper is organized as follows. Sec.\ref{sec:2} describes the theoretical background for the procedure of extracting of some additional information about the crystal structure by analyzing the variations of Patterson maps caused by the evolution of electron density of heavy atoms of the studied sample irradiated by the XFEL fs-pulse. In Sec.\ref{sec:3} we discuss the role of fluctuations that may appear due to the statistical dispersion of evolution paths and estimate the conditions under which the fluctuations can be neglected with sufficient degree of accuracy. In Sec.\ref{sec:4} we show the numerical results for the evolution of the electron density and separation of various forms of the Patterson maps for the arbitrary model crystal as a sample.

\section{Theoretical approach}
\label{sec:2}

In some extent our approach is analogous to the method of heavy atom in the X-ray diffraction analysis of proteins \cite{Vainshtein}. However, we do not assume that an additional heavy atom with known scattering factor is implanted artificially into the molecule under investigation. In order to explain the physical idea of our method let us consider the unit cell of the crystal that includes natural groups of relatively heavy atoms (for instance, sulfur S or iron Fe) and light atoms (for instance, oxygen O, nitrogen N, carbon C or hydrogen H). For matter of simplicity let us consider a unit cell with only two subsystems that consist of atoms of the same species: light atoms (L) and heavy atoms (H). In equilibrium state the scattering factor of such a unit cell can be written as \cite{Giac-Cryst}:

\begin{eqnarray}
\label{2-1}
    F (\vec q) = \sum^{N_L}_{L_i = 1} f_{L_i} (\vec q) e^{i \vec q \vec r_{L_i} } +
                \sum^{N_H}_{H_i = 1} f_{H_i} (\vec q) e^{i \vec q \vec r_{H_i} } =
                f_L (\vec q) S_L (\vec q) + f_H (\vec q) S_H (\vec q),
\end{eqnarray}

\noindent where $\vec q$ is the vector of momentum transfer, $\vec r_{L_i}$, $N_L$, $f_L$ and $\vec r_{H_i}$, $N_H$, $f_H$ represent the coordinates, total number and atomic scattering factors (ASF) of L- and H-atoms within the crystal unit cell correspondingly; $S_L$ and $S_H$ are the structure factors of the L- and H-subsystems which values depend on the positions of corresponding atoms within the crystal unit cell. They are defined as follows:

\begin{eqnarray}
\label{2-2}
    S_L (\vec q) = \sum^{N_L}_{L_i = 1} e^{i \vec q \vec r_{L_i} }, \quad S_H (\vec q) = \sum^{N_H}_{H_i = 1} e^{i \vec q \vec r_{H_i} }.
\end{eqnarray}

Here we assume that the energy of the XFEL pulse is far from the atomic shell ionization values and, as a result, the anomalous dispersion corrections in ASF can be omitted, so that it can be regarded as a real quantity.

In the framework of the kinematical theory the diffraction peak intensity is proportional to the square of absolute value of the form factor of the unit cell  \cite{Landau8}:

\begin{eqnarray}
\label{2-3}
    I_u (\vec q) = A \left| F ( \vec q) \right|^2 = A \left( f^2_L (\vec q) \left| S_L (\vec q) \right|^2 +
                    f^2_H (\vec q) \left| S_H (\vec q) \right|^2 +
                    2 f_L (\vec q) f_H (\vec q) \left| S_L (\vec q) \right| \left| S_H (\vec q) \right| \cos \Delta\varphi (\vec q) \right),
\end{eqnarray}

\noindent where $A$ is a constant which does not depend on the atomic parameters, $\Delta\varphi (\vec q) = \arg[S_L (\vec q)]-\arg[S_H (\vec q)]$ is the relative phase between the structure factors of both subsystems within the unit cell.

The Patterson function $P(\vec r)$ of the system can be obtained by inverse Fourier transform of the measured intensities (\ref{2-3}), \cite{Giac-Cryst,Patterson}. Following \cite{Giac-Cryst} this function represents a vector map of all possible combinations of atomic pairs peaking at points that correspond to all possible interatomic vectors. The height of these peaks is proportional to the product of ASF of atoms within such a pairwise combination. The total number of peaks is $N^2$ ($N$ is the total number of atoms within the crystal unit cell): $N$ peaks overlap at the origin of the cell and correspond to combination of every atom with itself), $N(N-1)$ are distributed within the unit cell and correspond to all other possible combinations of distances among the atoms. The main idea of the Patterson deconvolution consists in finding possible positions of heavy atoms and assuming some initial (trial) values for the phase angle of each reflection with respect to their basis. However, this task is often a tough challenge because some of these $N(N-1)$ peaks can overlap, and the highest peaks do not always correspond to heavy-heavy atom pairs (the problem becomes much more difficult with increasing $N$). The width of the Patterson peaks is also affected by the overlapping and is influenced by thermal smearing. Both leads to lower resolution of peak positions and makes a sensible interpretation of these peaks rather difficult. In order to improve the resolution within the Patterson map we suggest a method based on some specific features of interaction of XFEL femtosecond pulses with matter.

As it was shown earlier \cite{IUCRJ}, during the propagation of XFEL pulse through the crystal the ASF of the cell becomes dependent on specific parameters such as frequency $\omega$, duration $\tau$ and intensity $J$ due to the time evolution of the electron density:

\begin{eqnarray}
\label{2-4}
    F (\vec q) \rightarrow F (\vec q, \{ \kappa \} ) \Rightarrow I_u (\vec q) \rightarrow I_u (\vec q, \{ \kappa \} ),
\end{eqnarray}

\noindent where $\kappa$ represents a set of parameters of the XFEL pulse.

According to our calculations \cite{IUCRJ} it is possible to set specific experimental parameters so that the variation of ASF of H-atoms is relatively big whereas the ASF of L-atoms will change slightly:

\begin{eqnarray}
\label{2-5}
    \frac{f_L (\vec q, \{ \kappa \} )}{f_L (\vec q)} \equiv R_L (\vec q, \{ \kappa \} ) \approx 1, \quad
        \frac{f_H (\vec q, \{ \kappa \} )}{f_H (\vec q)} \equiv R_H (\vec q, \{ \kappa \} ) < 1.
\end{eqnarray}

The structure factors $S_L$ and $S_H$ remain unchanged because the atomic positions in a crystal are fixed in the fs time range. Hence, it is possible to derive the following expression for the diffraction intensities:

\begin{eqnarray}
\label{2-6}
    I_u (\vec q, \{ \kappa \} ) = A \left( f^2_L (\vec q) \left| S_L (\vec q) \right|^2 +
                    f^2_H (\vec q) R^2_H (\vec q, \{ \kappa \} ) \left| S_H (\vec q) \right|^2 +
                    2 f_L (\vec q) f_H (\vec q) R_H (\vec q, \{ \kappa \} )
                    \left| S_L (\vec q) \right| \left| S_H (\vec q) \right| \cos \Delta\varphi (\vec q) \right).
\end{eqnarray}

The values to be defined here are $\left| S_L (\vec q) \right|$, $\left| S_H (\vec q) \right|$ and $\cos \Delta\varphi (\vec q)$. Supposing one can use three different regimes of the pulse parameters corresponding to different sets of $\{\kappa \}$ and performs the corresponding diffraction experiments with noticeable contrast in the intensity distribution, then it will be possible to extract these unknown values solving the system of three equations having the form of (\ref{2-6}) (one should note that one of these regimes can be chosen in such a way that there is no evolution of ASF at all - for example, in a low-fluence mode). As a result, different Patterson functions for the L-L, L-H and H-H atomic combinations can be constructed which will enhance the initial resolution and will help to define the possible positions and phases of H-atom. A similar method of using several forms of Patterson maps extracted from the analysis of parametric X-ray radiation has been considered in \cite{PXR}.

The key step for implementing this scheme and checking the possibility of Patterson map separation consists in estimation of the $R_L$ and $R_H$ values. Besides, one should choose such a set of pulse parameters so that on the one hand condition (\ref{2-5}) could be fulfilled and on the other hand a reasonable contrast in intensity distribution functions should be obtained (otherwise the  system of equations of type (\ref{2-6}) becomes degenerated). The static values of ASF of L- and H-atoms can be analytically derived on the basis of effective charge model \cite{Fer-ECM-Acta,Fer-ECM-JAS}, dynamical values of ASF can be estimated on the basis of CEIX ("crystal evolution induced by X-ray") code described in \cite{IUCRJ} by simulation of evolution of electronic density of the crystal consisting of atoms of two different types and for various pulse parameters.

\section{The role of fluctuations}
\label{sec:3}

The major obstacle in solving the system of master equations of type (\ref{2-6}) is connected with possible fluctuations in atomic population probabilities of heavy atoms within the unit cell. Although we assume that evolution of every atom within the crystal unit cell is independent from each other, the variation of evolution paths could bring severe noise to the Patterson map making the interpretation rather difficult. Let us estimate the magnitude of such fluctuations and their influence on the shape of the diffraction pattern.

The most relevant quantity for the formation of the diffraction peak is the average value of ASF $\left\langle f(\vec q,t) \right\rangle$ describing the number of scattering electrons as a function of time and the momentum transfer $q = \frac{\sin \theta}{\lambda}$, where $\theta$ is the scattering angle and $\lambda$ is the photon wave length. This value becomes time-dependent due to the fast evolution of the electron density of atoms within the crystal unit cell. The statistical character of the ionization processes means that the ASF at the moment of time $t$ is a random value which depends on the probabilities of finding a certain electron configuration of the atom $P_{\lambda}(t)$ \cite{IUCRJ, Santra-Son}. Let us define the average ASF $\left\langle f(\vec q,t) \right\rangle$ and its standard deviation $\Delta f(\vec q,t)$ :

\begin{eqnarray}
\label{3-1}
    \left\langle f(\vec q,t) \right\rangle = \sum_{\lambda} P_{\lambda}(t) f_{\lambda}(\vec q), \nonumber\\
    \Delta f(\vec q,t) = \sqrt{\left\langle f^2(\vec q,t) \right\rangle - \left\langle f \right(\vec q,t) \rangle^2}, \\
    \left\langle f^2(\vec q,t) \right\rangle =  \sum_{\lambda} P_{\lambda}(t) f^2_{\lambda}(\vec q),\nonumber
\end{eqnarray}

\noindent where $f_{\lambda}(\vec q)$ is the stationary ASF value for the atomic configuration ${\lambda}$ at the momentum transfer $\vec q$. Since the anomalous dispersion term is omitted  we do not consider the energy range close to the resonance energy.

The calculation of the ASF value with probabilities $P_{\lambda}(t)$ related to one cell is performed by use of the ergodic hypothesis \cite{LandauV10} for a statistical ensemble of atoms within the whole crystal. It is also supposed that the fluctuations of  ASF for atoms in different cells are not correlated. In this case the ASF dispersion contributes only to the X-ray diffuse scattering background and does not change the intensity of the coherent diffraction peaks.

In the framework of the kinematical theory, the diffraction peak intensity is defined by the square of the ASF of all atoms within the crystal \cite{Landau8}. In the case of diffraction of XFEL pulse, the ASF values are fluctuating so that they should be averaged for all configurations:

\begin{eqnarray}
\label{3-2}
    I (\vec q) = \int^{+\infty}_{-\infty} \left\langle I (\vec q,t) \right\rangle J(t) dt, \nonumber\\
    \left\langle I (\vec q,t) \right\rangle= A \left\langle  \sum_a \sum_b f_a (\vec q,t)f_b^* (\vec q,t) e^{i \vec q (\vec R_a - \vec R_b)}  \right\rangle,
\end{eqnarray}

\noindent where $J(t)$ is the normalized envelope function of the XFEL pulse and the summation is performed over the coordinates $\vec R_a, \vec R_b$ of the atoms with ASF $f_a (\vec q,t)$ and $f_b (\vec q,t)$ in all unit cells of the crystal.

As it was mentioned above the average ASF and its mean-squared fluctuations are supposed to be the same for equivalent atoms in all unit cells and   fluctuations of ASF for different atoms are not correlated. In this case one can use the following equality:

\begin{eqnarray}
\label{3-3}
    \left\langle f_a (\vec q,t)f_b^* (\vec q,t)  \right\rangle  =  \left\langle f_a (\vec q,t)\right\rangle \left\langle f_b^* (\vec q,t)  \right\rangle +  \left[ \left\langle f_a (\vec q,t)f_b^* (\vec q,t)  \right\rangle - \left\langle f_a (\vec q,t)\right\rangle \left\langle f_b^* (\vec q,t)  \right\rangle \right] = \nonumber \\
    = \left\langle f_a (\vec q,t)\right\rangle \left\langle f_b^* (\vec q,t)  \right\rangle  +  \left[ \Delta f_a(\vec q ,t))^2 \right] \delta_{ab}.
\end{eqnarray}

Substituting (\ref{3-3}) into (\ref{3-2}) one can obtain:

\begin{eqnarray}
\label{3-4}
    \left\langle I (\vec q,t) \right\rangle = A \left\{ \left| \sum_a \left\langle f_a  (\vec q,t ) \right\rangle e^{i \vec q  \vec R_a  } \right|^2 +  \sum_a \left( \Delta  f_a(\vec q ,t) \right)^2 \right\}.
\end{eqnarray}

Summation in (\ref{3-4}) over the coordinates $\vec R_a$ of all cells in the crystal and over atoms within every unit cell is being performed by the following expansion for $\vec R_a$:

\begin{eqnarray}
\label{3-5}
    \vec R_a = \vec n + \vec r_j, \quad j = 1, 2,...,\sigma,
\end{eqnarray}

\noindent where $\vec n$ is the translation vector of the crystal lattice, and $\vec r_j$ and $\sigma$ are the coordinate of the $j$-th atom and the total number of atoms within the crystal unit cell correspondingly (translation symmetry of the crystal means that ASF depends on $j$ index only).

As a result, equation (\ref{3-4}) transforms into the following form:

\begin{eqnarray}
\label{3-6}
    \left\langle I (\vec q,t) \right\rangle = \left\langle I_u (\vec q,t) \right\rangle \left| \sum_{\vec n}e^{i\vec q \vec n} \right|^2 +
        A N \sum_j (\Delta  f_j(\vec q ,t))^2, \nonumber\\
    \left\langle I_u (\vec q,t) \right\rangle = A \left| \sum_j \left\langle f_j  (\vec q,t ) \right\rangle  e^{i\vec q \vec r_j} \right|^2,
\end{eqnarray}

\noindent where $\left\langle I_u (\vec q,t) \right\rangle$ is the total averaged diffraction intensity of the crystal unit cell.

The first term in (\ref{3-6}) defines the coherent diffraction intensity in accordance with the following identity \cite{solid}:

\begin{eqnarray}
\label{3-7}
    \left\langle I_{coh}(\vec q,t) \right\rangle =  \frac{N^2}{V}  \sum_{\vec h} \left\langle I_u (\vec h,t) \right\rangle \delta (\vec q - \vec h).
\end{eqnarray}

\noindent where $N$, $V$ and $\vec h$  are the total number of unit cells, the volume and the reciprocal lattice vectors of the crystal.

The second term in (\ref{3-6}) defines the background intensity of the diffuse (non-coherent) scattering. The relation of the total number of the diffuse photons scattered into the diffraction peak over the number of the coherent quanta scattered in the same angular interval is defined by the following parameter \cite{Landau8}:

\begin{eqnarray}
\label{3-8}
    \xi (t) \approx \frac{\lambda}{V^{1/3}} \frac{\sum_j (\Delta  F_j(\vec q ,t))^2}{\left\langle I_{coh}(\vec q,t) \right\rangle} \approx \frac{a}{L}\frac{\sum_j (\Delta  F_j(\vec q ,t))^2}{\left\langle I_{coh}(\vec q,t) \right\rangle},
\end{eqnarray}

\noindent where $L$ and $a$ are the linear dimensions of the crystal and the unit cell correspondingly (here one has taken into account that under diffraction conditions $\lambda_{max} \approx a$).

As it was shown in our paper \cite{IUCRJ}, the relative value of the standard deviation of fluctuation of ASF during the time of exposure of the crystal to the XFEL pulse does not exceed 20\%, so that even in the case of nanocrystals $\xi  \leq  0.01$ is expected. As a result, one can conclude that the second term in equation (\ref{3-6}) can be omitted remaining a high degree of accuracy so that the following approximate expression for the averaged intensity can be deduced:

\begin{eqnarray}
\label{3-9}
    \left\langle I (\vec q,t) \right\rangle = \left\langle I_u (\vec q,t) \right\rangle \left| \sum_{\vec n}e^{i\vec q \vec n} \right|^2.
\end{eqnarray}

Equation for $\left| S_L (\vec q) \right|$, $\left| S_H (\vec q) \right|$ and $\cos \Delta\varphi (\vec q)$ that is similar to (\ref{2-6}) can be obtained on the basis of (\ref{3-9}) by carrying out additional integration (\ref{3-2}) over time with pulse-weight function $J(t)$ and assuming that condition (\ref{2-5}) is fulfilled (i.e., the ASF value of L-atoms is time-independent):

\begin{eqnarray}
\label{3-10}
    \overline{\left\langle I (\vec q) \right\rangle } = \overline{\left\langle I_u
        (\vec q) \right\rangle} \left| \sum_{\vec n}e^{i\vec q \vec n} \right|^2,\nonumber\\
    \overline{\left\langle I_u (\vec q) \right\rangle } = A \left( f^2_L (\vec q) \left| S_L (\vec q) \right|^2 +
        \overline{\left\langle f^2_H (\vec q) \right\rangle} \left| S_H (\vec q) \right|^2 +
        2 f_L (\vec q) \overline{\left\langle f_H (\vec q) \right\rangle}
             \left| S_L (\vec q) \right| \left| S_H (\vec q) \right| \cos \Delta\varphi (\vec q) \right),
\end{eqnarray}

\noindent where $f_L (\vec q)$ is the static value of ASF of neutral (non-ionized) L-atoms, $\overline{\left\langle f_H (\vec q) \right\rangle}$ and $\overline{\left\langle f^2_H (\vec q) \right\rangle}$ are defined in the following way:

\begin{eqnarray}
\label{3-11}
    \overline{\left\langle f_H (\vec q) \right\rangle} = \int^{+\infty}_{-\infty} \left\langle f_H (\vec q,t) \right\rangle J(t) dt, \nonumber\\
    \overline{\left\langle f^2_H (\vec q) \right\rangle} = \int^{+\infty}_{-\infty} \left\langle f_H (\vec q,t) \right\rangle^2 J(t) dt.
\end{eqnarray}

One should note that both values of (\ref{3-11}) can be calculated by means of the CEIX code \cite{IUCRJ} with the same accuracy as ASF itself.

\section{Numerical results for atomic populations}
\label{sec:4}

\begin{figure}[t]
\includegraphics[scale=0.17]{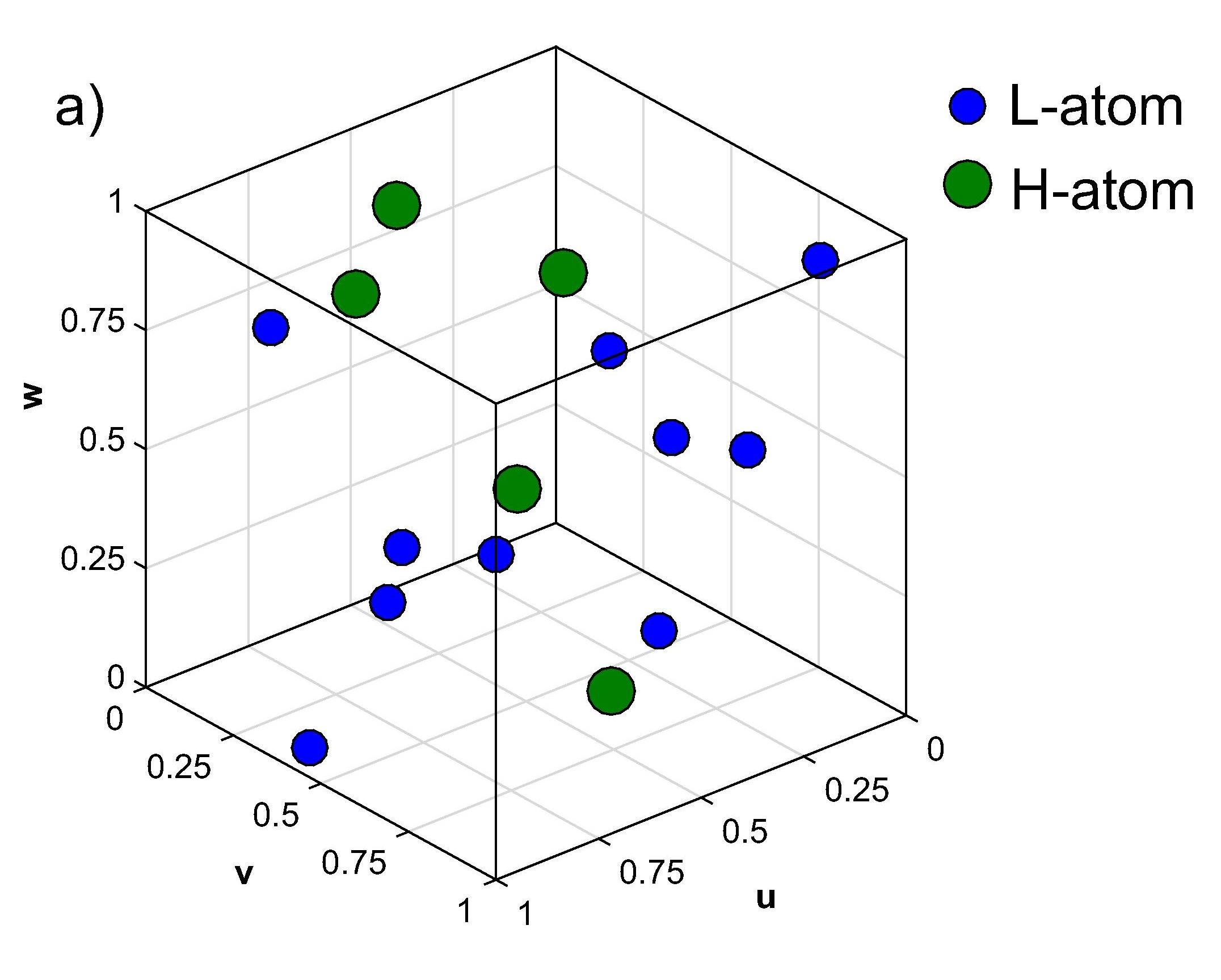}
\includegraphics[scale=0.17]{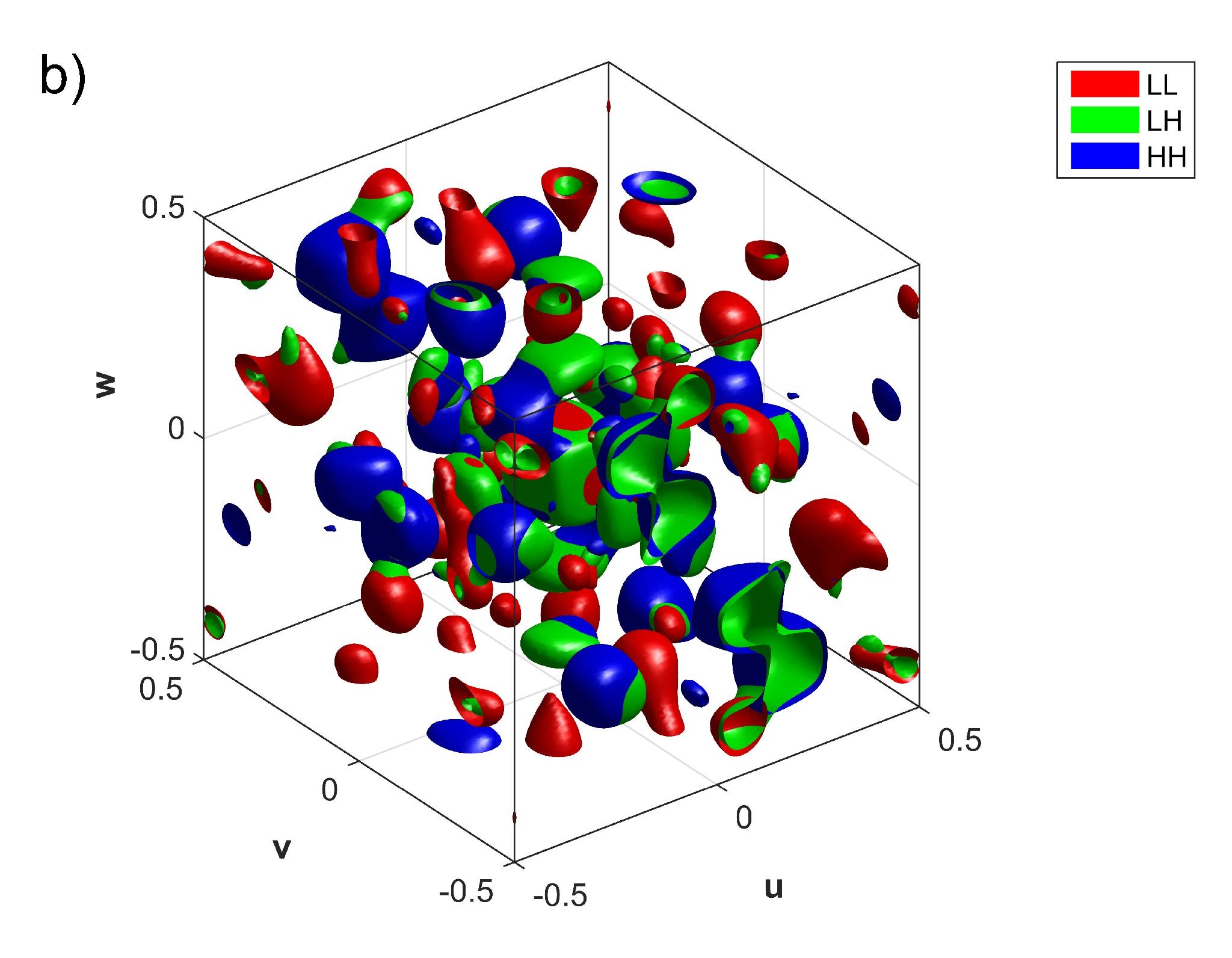}
\caption{The unit cell (a) and its Patterson map (b) of the model L-H crystal}
\label{Fig.1}
\end{figure}

In order to prove the applicability of the scheme (\ref{2-4})-(\ref{2-6}) (updated with (\ref{3-10})-(\ref{3-11}) ) and check the stability of the numerical algorithm for the cases of small contrasts an arbitrary L-H crystal with pre-defined positions of atoms was chosen as example (it is important to note that we tried to avoid any type of symmetry in the unit cell in order to investigate the general case of the model). In order to implement certain values of cross-sections and rates of the processes defining the evolution of the system the potassium and fluorine atoms have been selected as prototypes of H- and L-atoms correspondingly. The ratio of total electron charges of the constituting atoms is $Z_K/Z_F \approx 2.1$ that should be regarded as a valid condition for implementing the heavy-light atomic separation.
Fig.\ref{Fig.1} shows the real charge distribution function and the Patterson map (that was obtained numerically as a direct Fourier-transform of the charge distribution function) for the unit cell of model L-H crystal (the spacial linear scale  was chosen in the way so that the cell size is equal to unity): total number of L-atoms is 10, total number of H-atoms is 5.

\begin{figure}[t]
\includegraphics[scale=0.10]{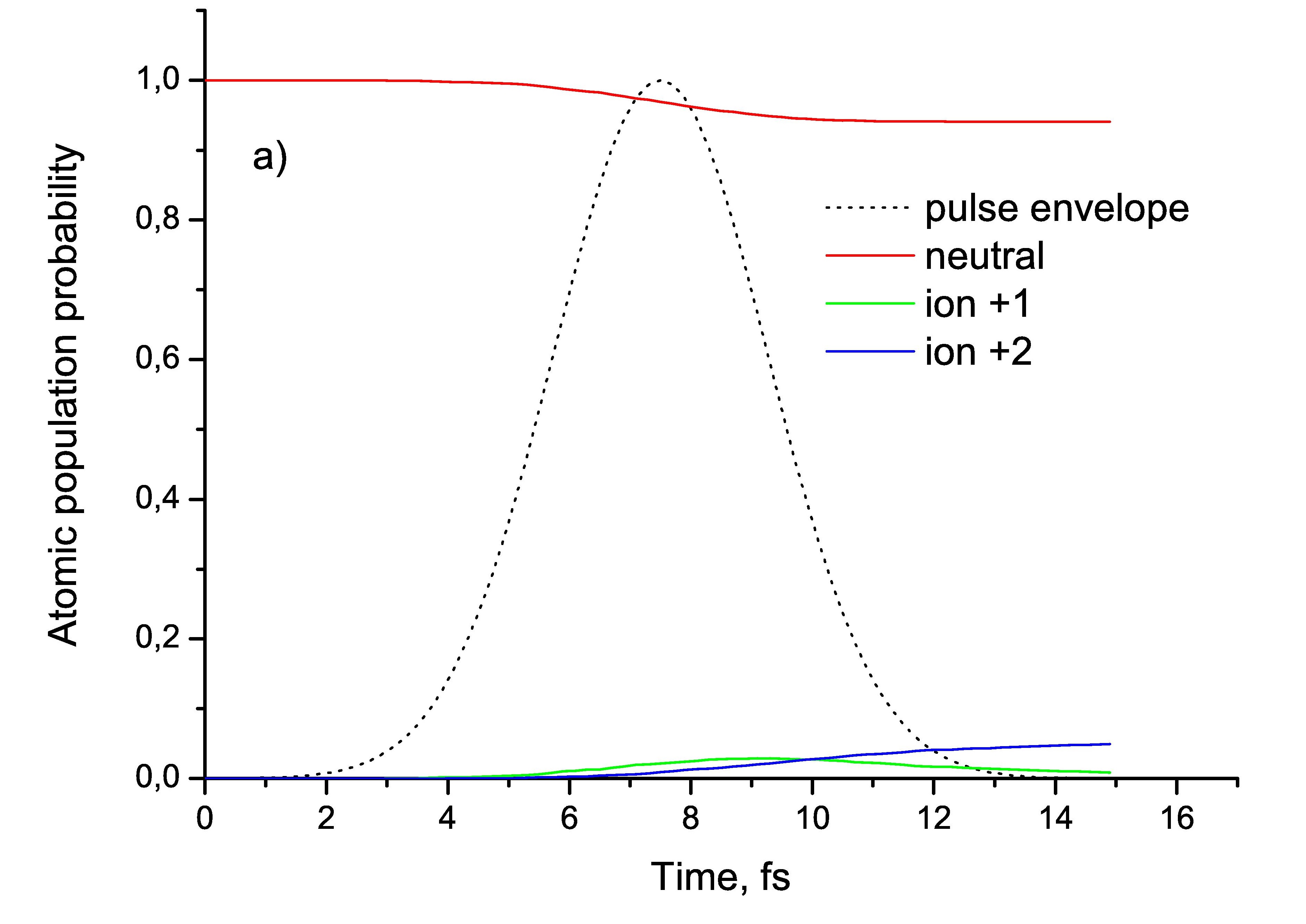}
\includegraphics[scale=0.10]{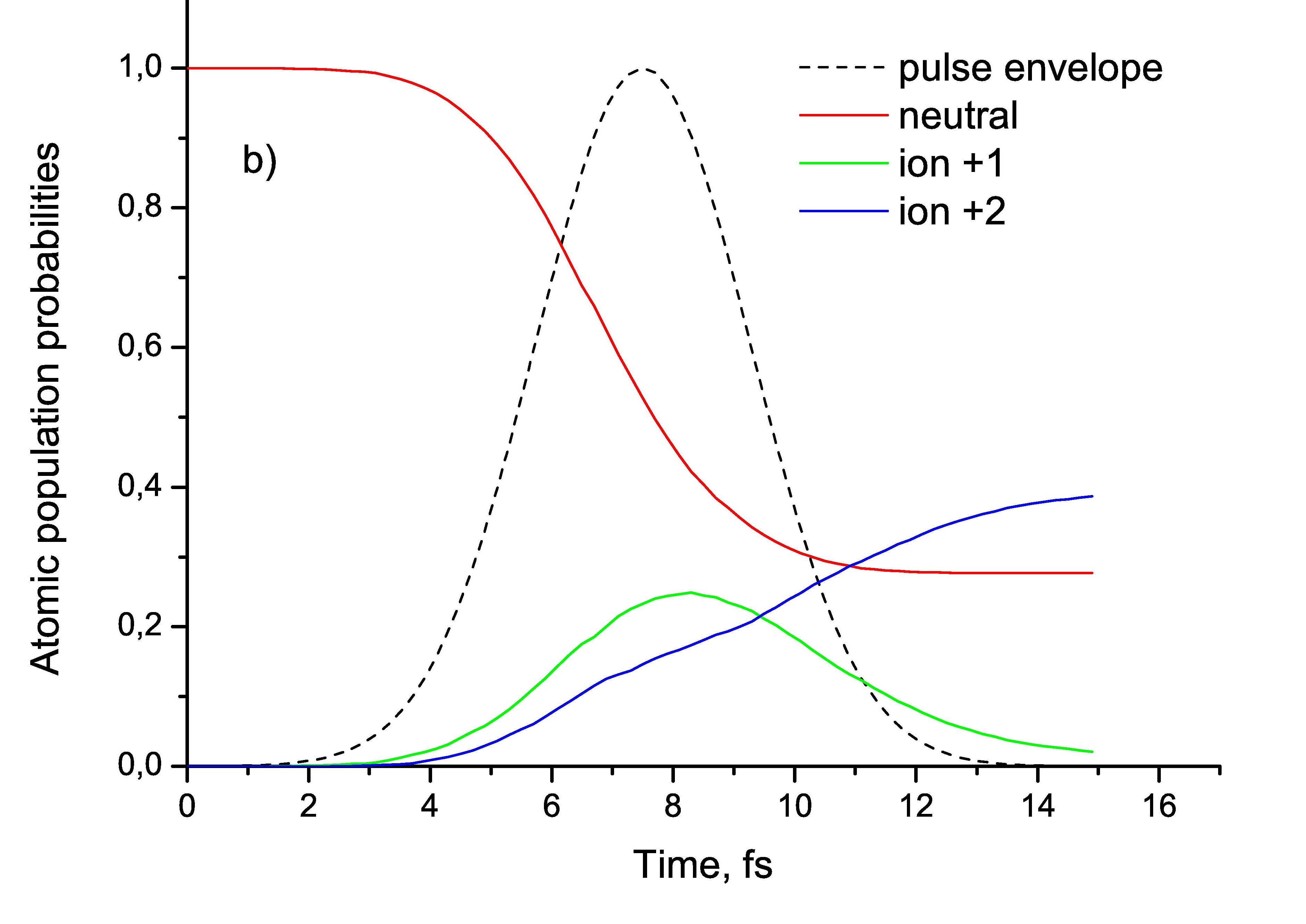}
\caption{Atomic population probabilities as a function of time: (a) L-atoms, (b) H-atoms.}
\label{Fig.2}
\end{figure}

Simulation of electron density evolution was performed by means of the CEIX code \cite{IUCRJ} with the following input conditions: the XFEL pulse was specified for a photon energy of 7 keV, Gaussian shape with 5 fs FWHM and a photon number of $10^{12}$ per pulse. Fig.\ref{Fig.2} shows an example of such a simulation regarding the distribution of atomic population probabilities as function of time (the beam focus spot was chosen to be 1 $\mu m^2$). As one can see, almost all L-atoms remain in the neutral state, whereas more than half of H-atoms are being ionized and the most populated state by the end of the pulse is the ion state with charge +2.

\begin{figure}[t]
\includegraphics[scale=0.10]{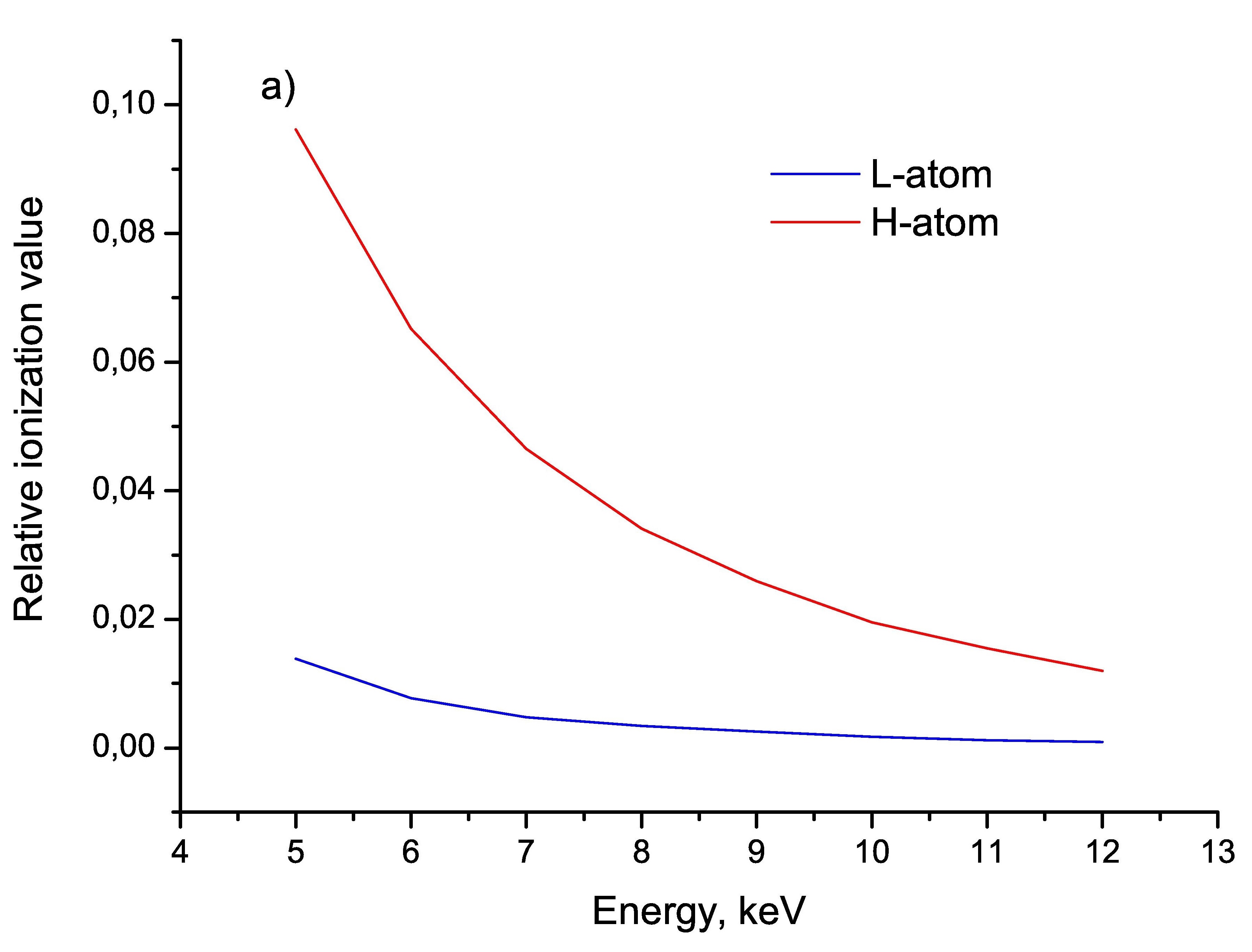}
\includegraphics[scale=0.10]{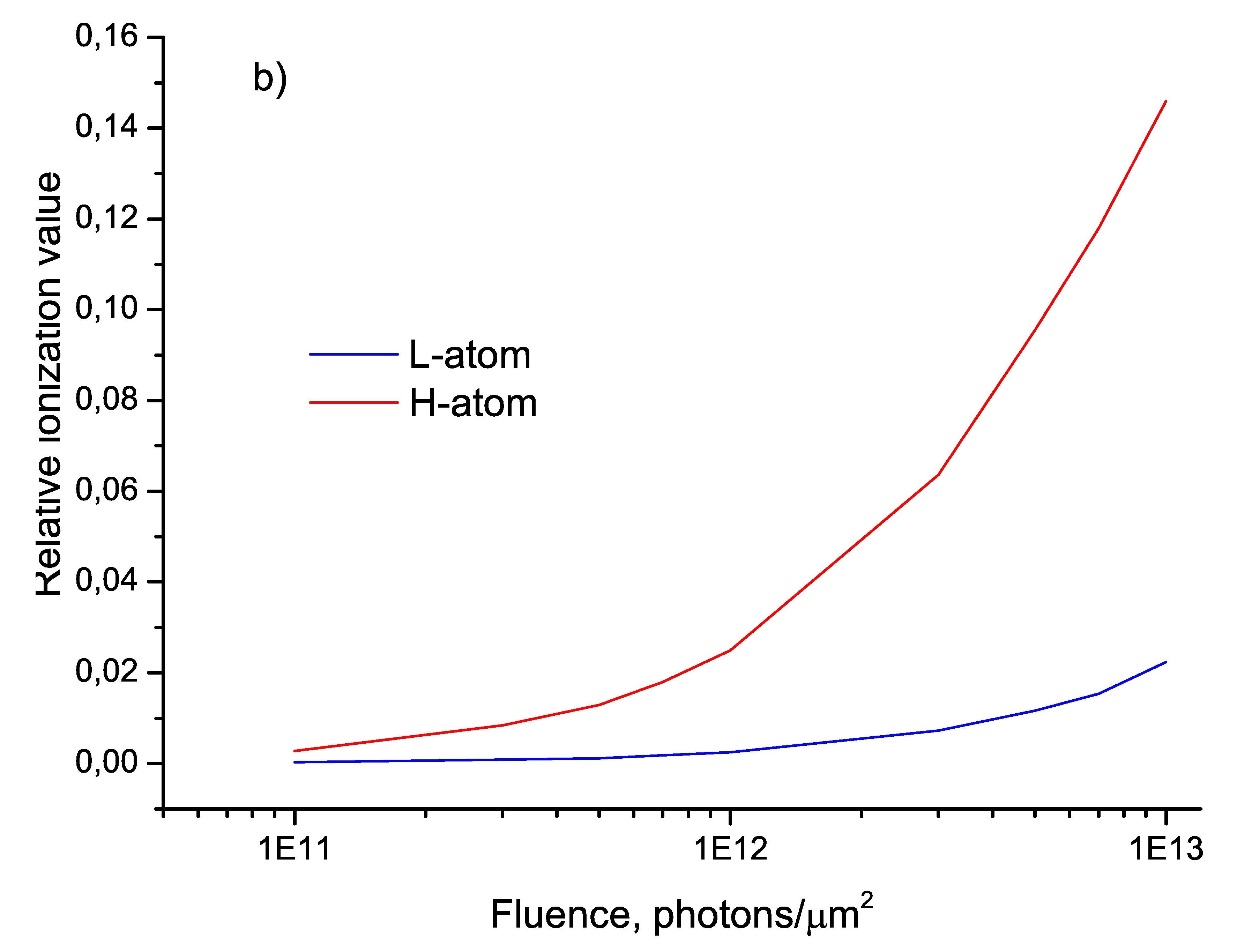}
\caption{Relative ionization value for L-atoms and H-atoms as a function of energy (a) and fluence (b).}
\label{Fig.3}
\end{figure}

In order to find different evolution regimes that could give non-degenerate contrast in the intensity distribution function the fluence of the pulse was subject to variation by changing the beam focus. Photoionization of the inner atomic shells is the main driving process that causes subsequent decay cascades and the main quantity responsible for its intensity is the photoionization cross-section value. For the selected prototypes of the constituting atoms at 7 kev the corresponding cross-sections for K over F differ by more than factor 20, so that condition (\ref{2-5}) was expected to be satisfied with high degree of accuracy. In order to check that we performed additional simulation to study the relative ionization value (the relative deviation of number of bound electrons in the ion from the number of bound electrons in the neutral state) as a function of incident photon energy (see Fig.\ref{Fig.3}a for details) at fluence of $10^{12}$ photons/$\mu m^2$. As one can see, at 7 keV there is almost no ionization in L-atoms, whereas 1-2 electrons are excited in every H-atom in average.

As it was mentioned above, the fluence was chosen to be the main parameter of ($\kappa$) variation. Fig.\ref{Fig.3}b shows the number of bound electrons per atom as function of fluence. One can conclude that within the considered range of fluences of $10^{11}-10^{13}$ photons/$\mu m^2$ our assumption regarding the absence of  a major evolution of electron density of  L-atoms remains valid, whereas for heavy atoms one can choose different evolution regimes that can define reasonable contrasts in the intensity distribution functions.

\begin{figure}[t]
\includegraphics[scale = 0.16]{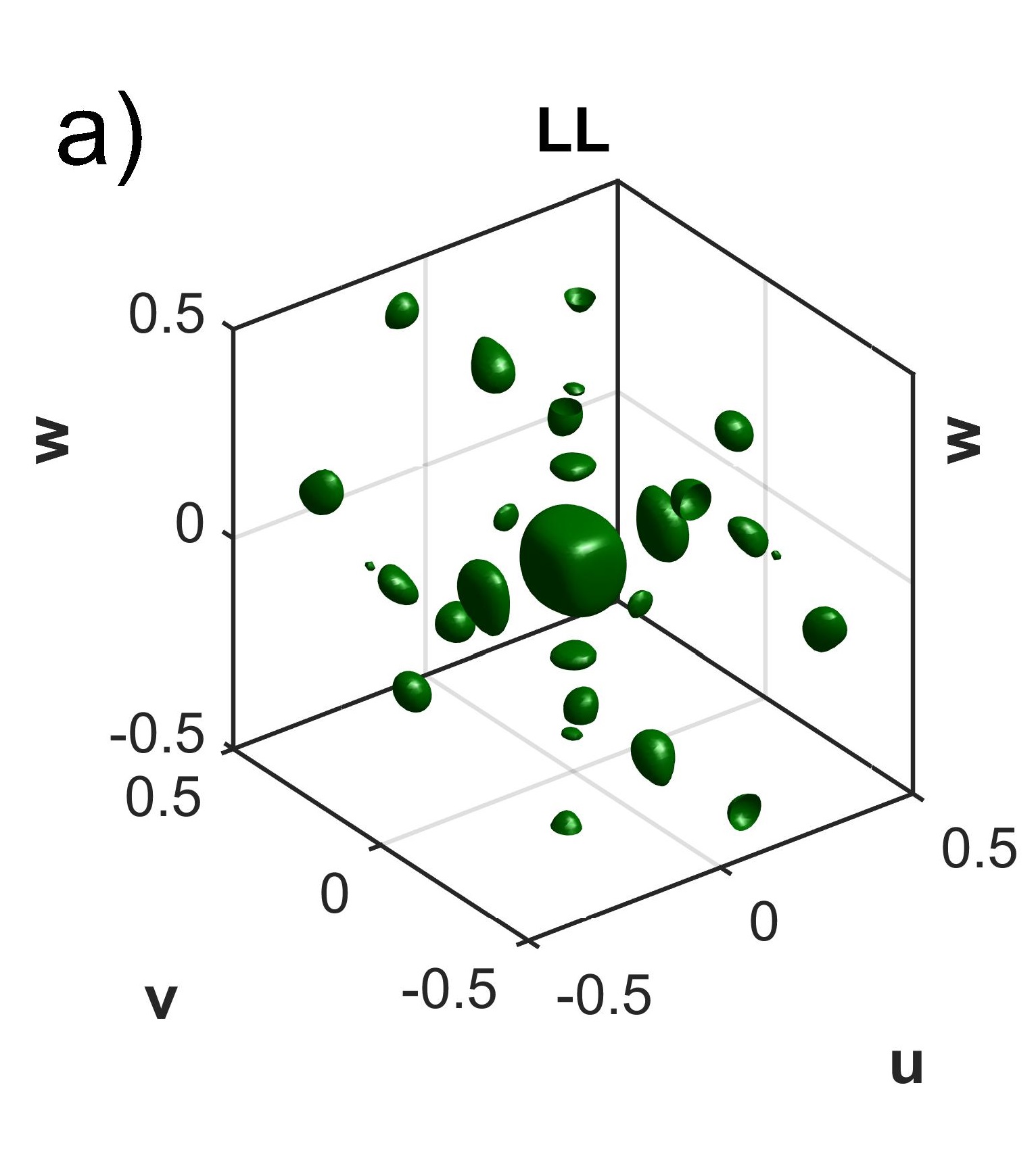}
\includegraphics[scale = 0.16]{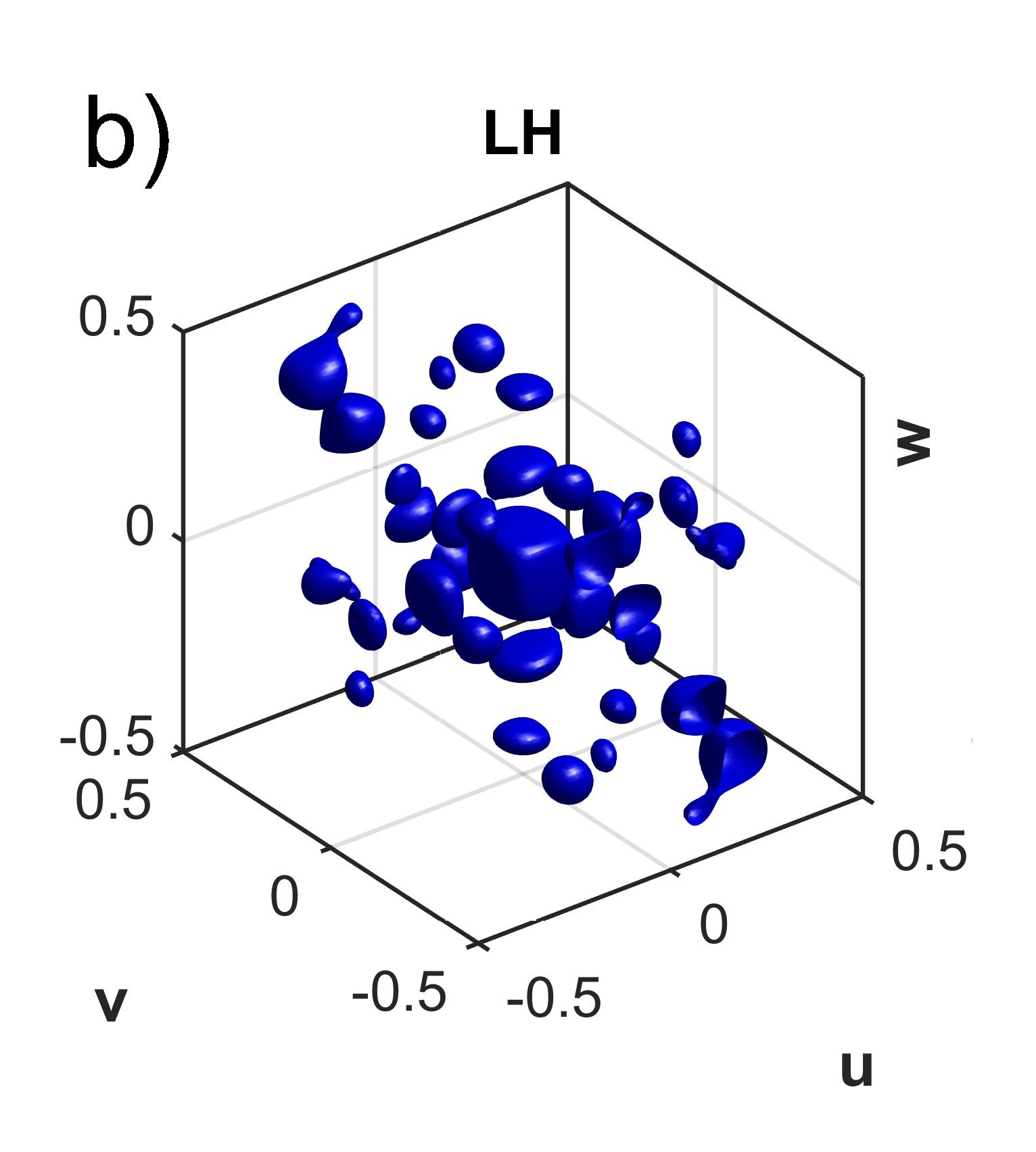}
\includegraphics[scale = 0.16]{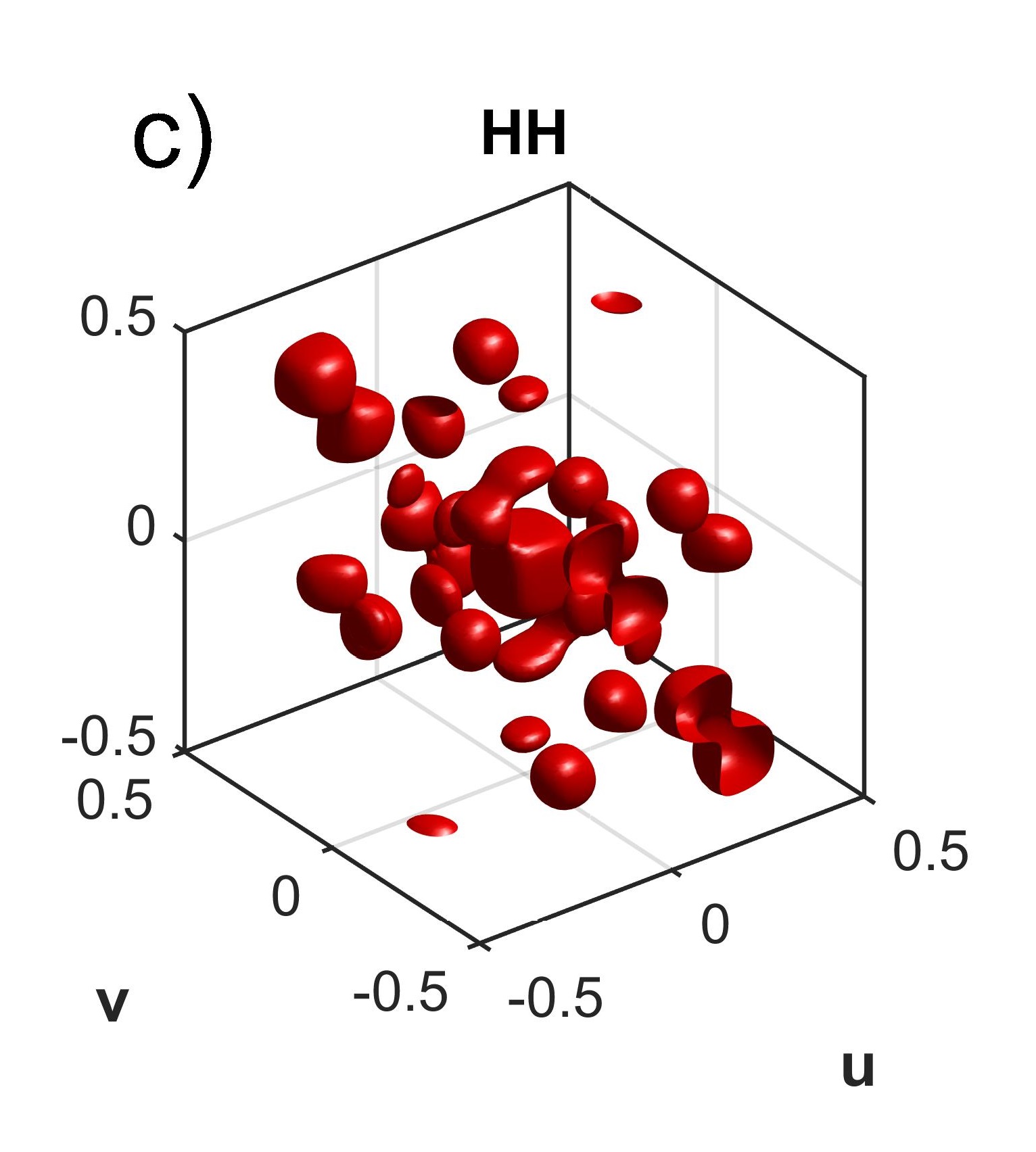}
\caption{Separate Patterson maps for the constituting atoms: (a) L-L atoms, (b) L-H atoms, (c) H-H atoms}
\label{Fig.4}
\end{figure}

In order to simulate intensity distribution functions on the basis of the charge distribution function we have chosen probe fluences with values of $0.5\times 10^{12}$ and $1.0\times 10^{12}$ photons/$\mu m^2$ and the corresponding table of 1331 reflections was constructed. Based on the obtained intensity distribution functions for 2 time-dependent and 1 time-independent case (low-fluence case that corresponds to the Patterson map shown on the Fig.\ref{Fig.1}b) a linear equation solver was implemented in order to calculate the absolute values of structure factors and the cosine of the relative phase between scattering factors of L- and H- atoms. As a result, the separate Patterson maps for L-L, L-H and H-H atomic combinations were constructed (see Fig.\ref{Fig.4}) and they coincided with the result of the direct calculation from the known charge distribution function that proved stability of the equation solver. One should note that this solution did not consider  possible fluctuations caused by the variation of some beam parameters like flux and energy.

\section{Discussion and conclusions}
\label{sec:concl}

The approach shown in the present paper enables one to create a separate Patterson map for the group of heavy atoms, the ASF of which is sensitive towards the radiation of XFEL fs-pulses. This additional information can be implemented in direct methods of finding the phases that are based on the phase retrieval procedure \cite{phase}. For instance, extracting the group of the strongest reflections is the starting point in the iteration techniques based on implementation of tangent formula \cite{tangent}. In the framework of the present approach such a group is extracted directly from the Patterson map of heavy atoms making implementation of the retrieval procedure easier than in the case of analysis of the full Patterson map of a macromolecule. It is important to stress that in the case of centrosymmetric crystals the present approach even allows one to clearly define the relative phase of the structure factors of light and heavy atoms (i.e., both the absolute value and its sign). Retrieval of the heavy atom positions and knowing this relative phase enables one to define positions of light atoms as well and reconstruct the total charge distribution function. In the case of non-centrosymmetric crystals the present approach allows one to define the absolute values of the relative phase and structure factors of both groups of atoms.

\section{Acknowlegements}

A. Leonov would like to thank the DAAD for financial support .


\begin{thebibliography}{[1]}

\bibitem{Taylor}
G. Taylor, Acta Cryst. Sec. D. \textbf{59}, 1881 (2003).

\bibitem{Giac-Phase}
C. Giacovazzo, Phasing in Crystallography, Oxford University Press, 2013.

\bibitem{Chapman2011}
H. Chapman et al., Nature \textbf{4}, 73 (2011).

\bibitem{FirstSLAC}
P. Emma et al., Nature Photon. \textbf{4}, 641 (2010).

\bibitem{FirstSACLA}
T. Ishikawa et al., Nature Photon. \textbf{6}, 540 (2012).

\bibitem{Redecke-SFX}
L. Redecke et al., Science \textbf{339}, 227 (2013).

\bibitem{Son-MAD}
S.-K. Son et al., Phys. Rev. Lett. \textbf{107}, 218102 (2011).

\bibitem{Spence-bB}
J.C. Spence et al., Opt. Express \textbf{19}, 2866 (2011).

\bibitem{Vainshtein}
B.K. Vainshtein, Modern Crystallography. Second Edition. Vol. 1, Oxford University Press, 2011.

\bibitem{Giac-Cryst}
C. Giacovazzo, Fundamentals  of Crystallography, Oxford University Press, 2011.

\bibitem{Landau8}
L.D. Landau and E.M. Lifshitz, Electrodynamics of Condensed Matter. Nauka, Moscow, 1982.

\bibitem{Patterson}
A.L. Patterson, Z. Krist. (A) \textbf{90}, 517 (1935).

\bibitem{IUCRJ}
A. Leonov, D. Ksenzov, A. Benediktovitch, I. Feranchuk and U. Pietsch, IUCrJ. \textbf{1}, 402–417, (2014).


\bibitem{PXR}
I.D.Feranchuk and A.P.Ulyanenkov, Acta Cryst. \textbf{A57}, 283 (2001).


\bibitem{Fer-ECM-Acta}
I.D. Feranchuk et al., Acta Cryst. A. \textbf{58}, 370 (2002).

\bibitem{Fer-ECM-JAS}
V.V. Triguk, I.D. Feranchuk, J. Appl. Spectroscopy \textbf{77}, 749 (2011).


\bibitem{Santra-Son}
S.-K. Son, L. Young, and R. Santra, Phys. Rev. A. \textbf{83}, 033402 (2011).


\bibitem{LandauV10}
L.D. Landau and E.M. Lifshitz, Physical Kinetics. Fizmatlit, Moscow, 2001.

\bibitem{solid}
J.M. Ziman, Principles of The Theory of Solids. Cambridge University Press, 1972.

\bibitem{phase}
R. P. Millane, Phase retrieval in crystallography and optics, J. Opt. Soc. Am. A. \textbf{7}, 394–411, (1990).

\bibitem{tangent}
J. Karle and H. Hauptman, Acta Crystallogr. \textbf{9}, 635-651 (1956).





\end{thebibliography}
\end{document}